\documentclass[pra,reprint,showpacs]{revtex4-1}
\usepackage{amsfonts}
\usepackage{float}
\usepackage{makeidx}
\usepackage{amsmath,amssymb,graphics,epsfig,color,times,bbm}
\usepackage{ulem}

\begin{document}

\title{Quantum enhanced optical phase estimation with a squeezed thermal
state}
\author{Juan Yu$^{1}$}
\author{Yue Qin$^{1}$}
\author{Jiliang Qin$^{1,2}$}
\author{Hong Wang$^{1}$}
\author{Zhihui Yan$^{1,2}$}
\email{zhyan@sxu.edu.cn}
\author{Xiaojun Jia$^{1,2}$}
\email{jiaxj@sxu.edu.cn}
\author{Kunchi Peng$^{1,2}$}
\affiliation{$^{1}$State Key Laboratory of Quantum Optics and Quantum Optics Devices,
Institute of Opto-Electronics, Shanxi University, Taiyuan, 030006, P. R.
China \\
$^{2}$Collaborative Innovation Center of Extreme Optics, Shanxi University,
Taiyuan 030006, P. R. China}

\begin{abstract}
Quantum phase estimation protocols can provide a measuring method of phase
shift with precision superior to standard quantum limit (SQL) due to the
application of a nonclassical state of light. A squeezed vacuum state, whose
variance in one quadrature is lower than the corresponding SQL, has been
pointed out a sensitive resource for quantum phase estimation and the
estimation accuracy is directly influenced by the properties of the squeezed
state. Here we detailedly analyze the influence of the purity and squeezing
level of the squeezed state on the accuracy of quantum phase estimation. The
maximum precision that can be achieved for a squeezed thermal state is
evaluated, and the experimental results are in agreement with the
theoretical analyses. It is also found that the width of the phase
estimation interval $\Delta \theta $ beyond SQL is correlated with the
purity of the squeezed state.
\end{abstract}

\maketitle

\section{Introduction}

The question for measuring the phase of light has been a subject of great
debate since the early work of Dirac \cite{dirac}. Due to the inexistence of
the phase Hermitian operator, the true value of phase cannot be directly
measured. A general method is to find an observable Hermitian operator
associated with phase, such as field- or intensity-based quantities by
interferometric devices \cite{vit,ber,wis1,ou,zhang,lum}, and then deduce
the phase indirectly according to the measurement results. This indirect
measurement process for the value of phase shift is called phase estimation.
The accuracy of usual phase estimation is limited by standard quantum limit
(SQL) because of the vacuum fluctuation of quantized electromagnetic field
\cite{boix}. Phase estimation is a powerful measurement strategy to perform
accurate measurements of various physical quantities including length,
velocity and displacements \cite{gio,brau}, and it is the heart of many
quantum enhanced metrology applications, such as improvement of time and
frequency standards \cite{udem,hink}, gravitational wave detection \cite%
{ligo,ander}, interferometry based on interacting systems \cite{dunn,rie},
quantum imaging \cite{lupo,tsang,gha}, atomic clock \cite{brave} and
magnetometry \cite{Dani,Nusran,maze,tay}.

Since Caves proposed that a quantum state can break the limit of shot noise
in 1981 \cite{caves}, many optical systems \cite%
{ban,kac,yon,ani,zixin,sesh,fang,higg,boto,xiang} have proved that a real
quantum state, for instance a squeezed state and an entangled state, can
greatly improve the accuracy of phase estimation with a given average photon
number \cite{alex,dina}. The accuracy of phase estimation is influenced by
the properties of the quantum state. In the basic principle of quantum
optics, the fluctuation added in one quadrature should be equal to that
reduced in its orthogonal quadrature for an ideal squeezed state of light.
However, a realistic squeezed state is difficult to be exactly pure
especially for high-level squeezed state due to the existence of extra noise
in its generation system \cite{yang}. Accordingly, it is quite necessary to
explore the effect of the properties of the squeezed state on the phase
estimation results.

The theory of quantum phase estimation provides the ultimate bound on
precision of phase estimation in the form of quantum Cram\'{e}r-Rao bound
(QCRB), which is independent with detection strategies \cite{dong}. The QCRB
is essentially determined by Heisenberg uncertainty and is given by the
inverse of quantum Fisher information (QFI) associated with the resource.
The theoretical analyses show that homodyne measurement is optimal for
squeezed pure state but not optimal for squeezed thermal state, and the
maximum precision that can be achieved for squeezed thermal state via
homodyne measurement is called optimal Cram\'{e}r-Rao bound (OCRB) \cite{asp}%
. In Ref. \cite{berni}, a squeezed-enhanced phase estimation is realized
with the help of feedback control. Here, we detailedly analyze the influence
of properties of squeezed state on the phase estimation results. By using a
squeezed thermal state as the probe beam, the effects of the squeezing level
and the purity of a squeezed state on the phase estimation results are
given. Then, we experimentally implement that the absolute phase estimation
can be enhanced with much higher squeezing level and squeezed pure state
behaves the optimal resource to reach QCRB. Our research is of general
interest in the sense of phase estimation based on the squeezing mechanism
and the results provide a reference for multi phase estimation based on
multipartite entanglement \cite{Proie,Zhangc}.

\section{Phase estimation with a squeezed thermal state}

An optical field can be represented by the annihilation operator $\hat{a}$
in quantum mechanics. The orthogonal amplitude and phase operators can be
represented in terms of the creation and annihilation operators as $\hat{x}%
=\left( \hat{a}+\hat{a}^{\dagger }\right) /\sqrt{2}$ and $\hat{p}=i\left(
\hat{a}^{\dagger }-\hat{a}\right) /\sqrt{2}$, $\hat{x}$ and $\hat{p}$
satisfy the commutation relations $\left[ \hat{x},\hat{p}\right] =i$.
Usually, a coherent state or a vacuum state is a minimum uncertainty state
and the variances of the two quadrature components are equal: $\langle
\Delta ^{2}\hat{x}\rangle =\langle \Delta ^{2}\hat{p}\rangle =1/2$. A
squeezed state is defined as its variance of one quadrature is reduced
relative to the corresponding SQL while the variance of its orthogonal
quadrature is amplified. The squeezing parameter $r$ is used to indicate the
squeezing level of the squeezed state, i. e. the variance in a squeezed
quadrature, $e^{-2r}/2$, is always below the corresponding SQL \cite{huo1}.
The mean photon number of pure squeezed vacuum state is $n=\sinh ^{2}r$. In
the past decades, squeezed states of light have been obtained by several
groups and squeezing level has been improved continually \cite%
{meh,yang1,vahl,jian}. In the actual experimental generation processing,
there is some inevitably extra noise in its antisqueezing quadrature
component. A extra antisqueezing parameter $r^{\prime }$ is introduced to
describe the extra noise level of antisqueezing quadrature component \cite%
{zhou} and the covariance matrix of this squeezed state is expressed as $%
\mathbf{\sigma }_{0}=1/2$ Diag$(e^{-2r},e^{2r+2r^{\prime }})$, which is
usually called a squeezed thermal state. The mean photon number of squeezed
thermal state is $n=e^{r^{\prime }}n_{r}+(e^{r^{\prime }}-1)/2$, where $%
n_{r}=\sinh ^{2}(r+r^{\prime }/2)$ is the photon number contributing from
the squeezing effect, $(e^{r^{\prime }}-1)/2$ is the photon number
contributing from the extra noise in antisqueezing quadrature \cite{asp}.

\begin{figure}[tbph]
\centerline{\includegraphics[width=0.75\columnwidth]{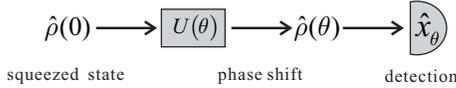}}
\caption{Schematic of a general quantum phase estimation. A squeezed state $%
\hat{\protect\rho}\left( 0\right) $ undergoes an unknown phase shift $%
\protect\theta $. A function of the data samples $\hat{x}_{_{\protect\theta %
}}$ associated with the phase shift are measured by a detection strategy.}
\end{figure}

A general scheme of a quantum phase estimation with a squeezed state is
shown in Fig. 1. A squeezed state $\hat{\rho}\left( 0\right) $ undergoes a
phase shift described by a unitary operator $\hat{U}\left( \theta \right)
=\exp \left( -i\theta \hat{n}\right) $,
\begin{equation}
\hat{\rho}\left( \theta \right) =\hat{U}\left( \theta \right) \hat{\rho}%
\left( 0\right) \hat{U}^{\dagger }\left( \theta \right) ,
\end{equation}%
where $\hat{n}=\hat{a}^{\dagger }\hat{a}$ is a number operator and $\theta $
is the phase shift to be estimated. The output state is detected by a
detection strategy and the obtained data samples associated with the phase
shift are processed by Bayesian inference \cite{wie}. The essence of phase
shift operation is a rotation operation on an initial state, and the
quadrature operator is expressed as $\hat{x}_{_{\theta }}=\left( \hat{a}%
e^{-i\theta }+\hat{a}^{\dagger }e^{i\theta }\right) /\sqrt{2}$. The variance
of the squeezed state $\hat{\rho}\left( \theta \right) $ is associated with
the phase shift acted on the probe beam, and the phase shift can be
indirectly obtained by measuring the variance of quadrature $\hat{x}_{\theta
}$ via homodyne detection. Thus, the phase estimation protocol we provided
here is only appropriate for a squeezed state. In general, the variance of
phase shift \text{Var}$\left[ \theta \right] $ for any unbiased estimator is
bounded at the times of measurements $N$ by the Cram\'{e}r-Rao theorem \cite%
{cra}:

\begin{equation}
\text{Var}\left[ \theta \right] \geq \frac{1}{NF\left( \theta \right) },
\end{equation}%
where $F\left( \theta \right) $ is the Fisher information (FI) \cite{hass},
which is the observed information about the unknown parameter. The quantum
Fisher information (QFI) $H$ is the maximized FI over all possible detection
schemes, i. e. $F\left( \theta \right) \leq H$. According to Ref. \cite{gao}%
, $H$ can be fully expressed in terms of the covariance matrix of the
Gaussian state and the QCRB of the quantum phase estimation for a
single-mode squeezed thermal state is:

\begin{equation}
\text{Var}_{\text{sq}}\left[ \theta \right] =\frac{1}{8Nn_{r}\left(
n_{r}+1\right) }\left[ \frac{1}{2}+\frac{1}{2}e^{-2r^{\prime }}\right] ,
\end{equation}%
Comparing with a coherent state to be used as probe beam in phase estimation
($\sim $ $1/\left( 4Nn\right) $, where $n$ is the mean photon number of
probe beam) \cite{asp}, the estimation accuracy can be greatly improved with
the help of a squeezed state. For a squeezed pure state, there is not any
extra noise in the antisqueezing quadrature component, i. e. $r^{\prime }=0$%
, and the QCRB becomes Var$\left[ \theta \right] =1/[8Nn_{r}\left(
n_{r}+1\right) ]$ \cite{alex,rob,geno}.

\section{Reachable bound with homodyne detection and Bayesian inference}

Homodyne detection is a common detection strategy for state reconstruction
in continuous-variable (CV) regime \cite{gran}. It is a kind of simple and
accurate detection strategy because it can provide a phase reference for
estimating the value of the phase shift.\textit{\ }The data samples $\{\hat{x%
}_{\theta }\}$ associated with the phase shift $\theta $ are obtained
through a local projective Von Neuman measurement and then the true value of
phase shift can be indirectly deduced according to the measurement results
\cite{huo}.

In order to evaluate the maximum precision that is achieved for a squeezed
thermal state with homodyne measurement, we use the Wigner function to
describe our system \cite{pinel,oliv,chen}. In the Wigner function
description, the quadratures of the probe beam correspond to two phase-space
coordinates $x$ and $p$, which can be grouped into a two-dimensional vector $%
\mathbf{X}$, $\mathbf{X}^{T}=(x,p)$. The Wigner function associated with the
shifted squeezed thermal state $\hat{\rho}_{_{\beta ,r}}\left( \theta
\right) $ is:
\begin{equation}
W_{\theta }\left( \mathbf{X}\right) =\frac{\exp [-\frac{1}{2}\mathbf{X}^{T}%
\mathbf{\sigma }_{\theta }^{-1}\mathbf{X}]}{2\pi \sqrt{\text{Det}[\mathbf{%
\sigma }_{\theta }]}},
\end{equation}%
where $\mathbf{\sigma }_{\theta }$ is the covariance matrix after the phase
shift,
\begin{widetext}
\begin{equation}
\sigma _{\theta }=\frac{1}{2}\left(
\begin{array}{cc}
e^{-2r}\cos ^{2}\theta +e^{2r+2r^{\prime }}\sin ^{2}\theta  & \frac{1}{2}%
(e^{2r+2r^{\prime }}-e^{-2r})\sin \left( 2\theta \right)  \\
\frac{1}{2}(e^{2r+2r^{\prime }}-e^{-2r})\sin \left( 2\theta \right)  &
e^{2r+2r^{\prime }}\cos ^{2}\theta +e^{-2r}\sin ^{2}\theta
\end{array}
\right) .
\end{equation}
\end{widetext}Then the individual marginal probability distribution $p\left(
x|\theta \right) $ conditioned on single homodyne measurement outcome of a
shifted squeezed thermal state is calculated from the Wigner function \cite%
{leon}:

\begin{equation}
p\left( x|\theta \right) =\int_{\mathbb{R}}W_{\theta }\left( \mathbf{X}%
\right) dy=\frac{1}{e^{r^{\prime }}\sqrt{\pi \Sigma _{\theta }^{2}}}\exp [-%
\frac{x_{\theta }^{2}}{e^{2r^{\prime }}\Sigma _{\theta }^{2}}],
\end{equation}%
where $\Sigma _{\theta }^{2}=[e^{-2r-2r^{\prime }}\cos ^{2}\theta
+e^{2r}\sin ^{2}\theta ]$ is the variance of the probe beam, $\{x_{\theta
}\} $ is the noise distribution of the squeezed state associated with $%
\theta $ obtained from the homodyne measurement. The FI can be easily
evaluated from its definition \cite{paris}:

\begin{eqnarray}
F\left( \theta \right) &=&\int_{\mathbb{R}}p\left( x|\theta \right) \left[
\frac{\partial \log p\left( x|\theta \right) }{\partial \theta }\right]
^{2}dx  \notag \\
&=&\frac{\sin ^{2}\left( 2\theta \right) \left( e^{2r}-e^{-2r-2r^{\prime
}}\right) ^{2}}{2\left( \Sigma _{\theta }^{2}\right) ^{2}}.
\end{eqnarray}%
It is obvious that the expression of the FI is dependent on the phase shift $%
\theta $ and the squeezing parameters $r$ as well as the extra antisqueezing
parameter $r^{\prime }$. The maximum of the FI can be achieved at an optimal
phase $\theta _{\text{opt}}=1/2\arccos (\tanh (2r+r^{\prime }))$ and $%
F_{\max }$ with homodyne measurement is,

\begin{equation}
F_{\max }=2\sinh ^{2}(2r+r^{\prime }).
\end{equation}

Then upon using the Cram\'{e}r-Rao theorem, the variance of optimal phase
estimation with homodyne measurement goes as:

\begin{equation}
\text{Var}_{\text{sq}}^{\hom }\left[ \theta \right] =\frac{1}{%
8Nn_{r}(n_{r}+1)}.
\end{equation}
It means that the homodyne measurement is not optimal for a squeezed thermal
state by comparing Eq. (3) and (9) and the estimation accuracy can attain
the optimal Cram\'{e}r-Rao bound (OCRB) \cite{berni}.

Bayesian inference, which is known as \textquotedblleft probability
theory\textquotedblright, is the theory of how to combine uncertain
information from multiple sources to make optimal decision under
uncertainty. If $x$ is the variable associated with the phase shift, then
the Bayes' rule states:

\begin{equation}
p\left( \theta |x\right) =\frac{p\left( x|\theta \right) p\left( \theta
\right) }{p\left( x\right) },
\end{equation}%
where $p\left( \cdot |\cdot \right) $ are the conditional probabilities
about parameters $x$ and $\theta $. $p\left( x|\theta \right) $ is the
marginal probability distribution of the shifted squeezed thermal state and $%
p\left( \theta |x\right) $ is the posteriori probability distribution (PPD)
of the phase shift. $p\left( x\right) $ are the total probabilities to
observe $x$ and $p\left( \theta \right) =2/\pi $ is the prior information
which is a flat distribution. The result of each measurement is used as a
prior information for the next measurement. The PPD $p\left( \theta
|x\right) $ based on $N$ sampled homodyne measurements is given by:

\begin{equation}
p\left( \theta |x\right) =\frac{1}{\mathcal{N}}\underset{k=1}{\overset{N}{%
\prod }}p\left( x_{_{k}}|\theta \right) ,
\end{equation}%
where $\mathcal{N=}\int_{0}^{\frac{\pi }{2}}p\left( \theta |x\right) d\theta
$ is a normalization constant, $p\left( x_{_{k}}|\theta \right) $ is the
individual marginal probability distribution conditioned on each homodyne
measurement which are given by Eq. (6).

\section{Experimental setup and result}

\begin{figure}[tbph]
\centerline{\includegraphics[width=0.75\columnwidth]{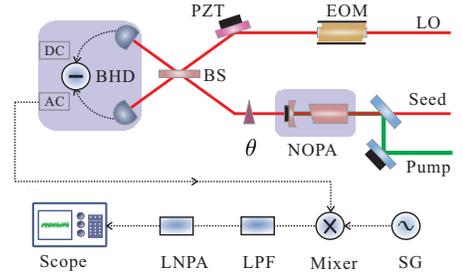}}
\caption{Schematic of the experimental setup. Dotted curves and solid curves
represent the circuitry part and the light path, respectively. NOPA,
nondegenerate optical parametric amplifier; Pump, pump field of the NOPA;
Seed, seed field of the NOPA; LO, a strong local oscillator beam; EOM,
electro-optic modulator; PZT, piezo-electric transducer; BS, 50/50 beam
splitter; BHD, balanced homodyne detector; DC, direct-current signal; AC,
alternating-current signal; SG, signal generator; $\protect\theta $,
electric phase controller; LPF, low-pass filter; LNPA, low noise
pre-amplifier; Scope, oscilloscope.}
\end{figure}

A schematic of experimental setup is illustrated in Fig. 2, which includes a
source of squeezed state, a balanced homodyne detection system, a phase
control system and a data acquisition system. A squeezed state is produced
by a nondegenerate optical parametric amplifier (NOPA), which is pumped by a
continuous wave intra-cavity frequency-doubled tunable single-frequency
Nd:YAP/LBO solid-state laser provided by YuGuang company CDPSSFG-VIB (not
shown in Fig. 2). The output fundamental wave at wavelength of 1080 nm is
used for the injected seed beams of NOPA and the local oscillator beam of
homodyne detection system. The second harmonic wave at 540 nm serves as the
pump field of the NOPA. The NOPA consists of an $\alpha $-cut KTP crystal
and a concave mirror, which can realize type-II non-critical phase matching
without walk-off effect. The front face of the crystal is highreflection
(HR) coated for 1080 nm and T$_{\text{1}}$ = 18\% coated for 540 nm, which
serves as the input coupler. The end face of the KTP is cut to 1$^{\circ }$
along y-z plane of the crystal and is antireflection coated for both 1080 nm
and 540 nm. The concave mirror with a radius of curvature of 50 mm coated
with T$_{\text{2}}$ = 12.5\% for 1080 nm and HR for 540 nm serves as the
output coupler, which is mounted on a piezoelectric transducer to actively
lock the cavity length of NOPA on resonance with the injected signal at 1080
nm. Through an intracavity frequency down conversion processing in the NOPA,
an Einstein-Podolsky-Rosen (EPR) entangled state of light or two single-mode
squeezed states of light at 1080 nm with orthogonal polarizations can be
generated separately \cite{zhou1}. The squeezed state with different
squeezing parameter $r$ and different purity can been generated by
controlling the experimental conditions.

The generated squeezed state acquires an unknown phase shift $\theta $
within a range of $[0,$ $\pi /2]$ and then is combined with a strong local
oscillator beam (5 mW) at a 50/50 beam splitter (BS) for homodyne
measurement. The relative phase control between the local oscillator beam
and probe bem is achieved by an improved Pound-Drever-Hall (PDH) technique
\cite{deng}. The local oscillator beam is phase-modulated by an
electro-optic modulator (EOM) with a sine signal at 7.3 MHz. The first-step
error signal is obtained by mixing alternating-current (AC) signal detected
by the homodyne detector and the sine signal modulated on the EOM. The final
error signal to realize the phase locking of probe beam and the local
oscillator beam to a specific degree is obtained by coupling the first-step
error signal with the direct-current (DC) output from the homodyne detector
by a certain percentage. Finally, the error signal is feedback to the
piezo-electric transducer (PZT) attached on a high reflection mirror. The
experimental data $\{x_{\theta }\}$ of quantum phase estimation is recorded
by an oscilloscope via quantum tomography technique. A PPD of $\theta $
conditioned on the $N$ sampled homodyne measurements can be calculated
according to Eq. (11).

\begin{figure}[tbph]
\centerline{\includegraphics[width=0.75\columnwidth]{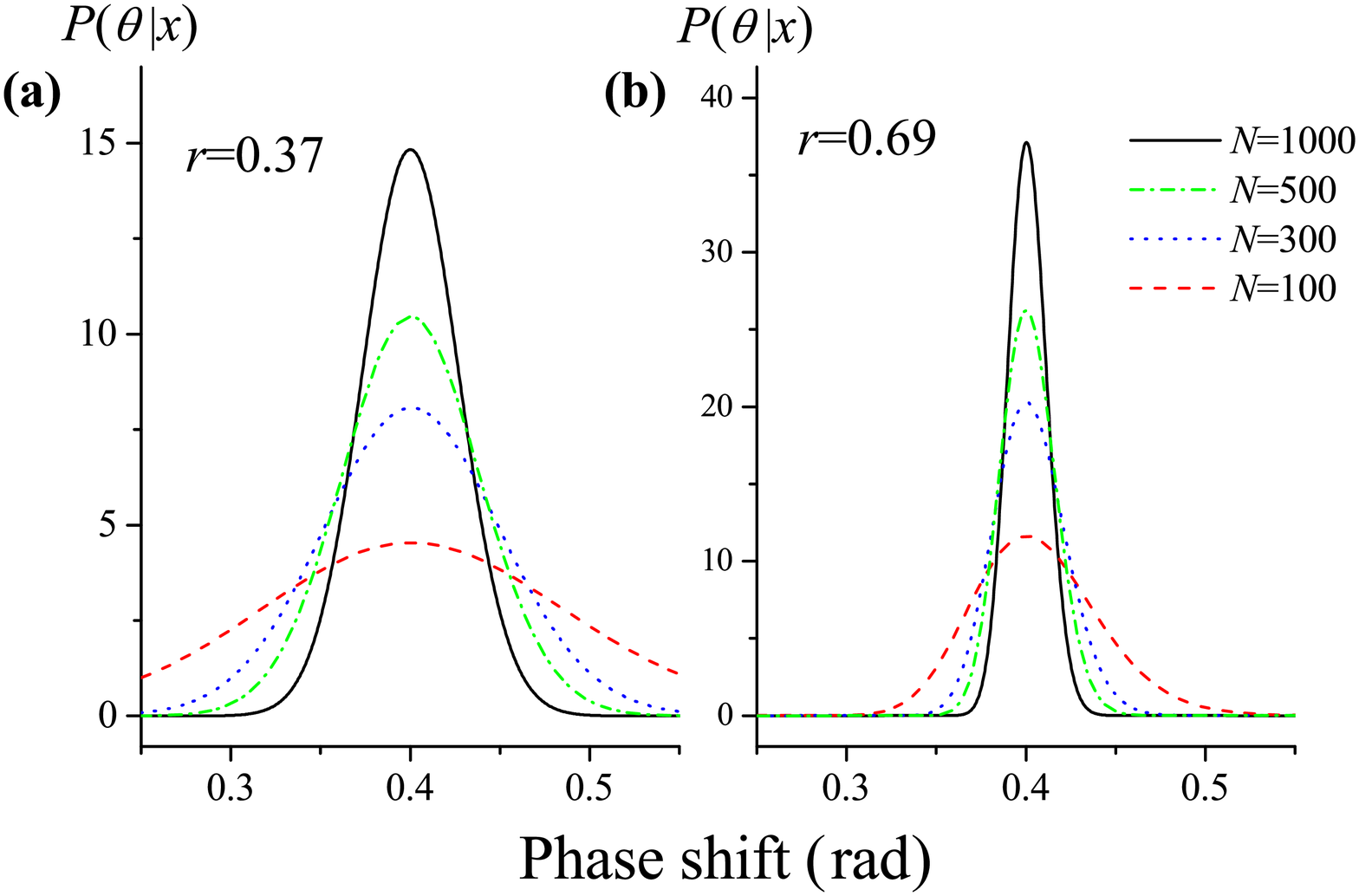}}
\caption{Posteriori probability distributions for different values of the
involved parameters at a fixed phase shift $\protect\theta=0.4 $. Fig. 3 (a)
and 3 (b) show PPDs versu phase shift for different numbers of homodyne
samples $N $ when the squeezing parameter $r$ of 0.37 and 0.69,
respectively. }
\end{figure}

To investigate the performance for different times of measurements $N$ and
squeezing parameter $r$, we fix the phase shift at $\theta =0.4$ firstly in
the experiment. The PPDs of the phase shift conditioned on the sampled
homodyne data are obtained for different values of the involved parameters
as a function of $\theta $ as shown in Fig. 3. The solid black, dash dot
green, dot blue and dash red curves correspond to $N=$ 1000, 500, 300, 100,
respectively. A suitable estimator for the actual value of a fixed phase
shift is given by the maximum of the distribution because of the symmetric
form of the PPD. The definition of the variance is Var$\left[ \theta \right]
=\left\langle \theta ^{2}\right\rangle -\left\langle \theta \right\rangle
^{2}$, and the Var$\left[ \theta \right] $ is given by $1/NF(\theta )$,
which is calculated from the homodyne measurements. Phase estimation can be
enhanced with much higher squeezing parameter $r$ and more times of
measurements $N$.

\begin{figure}[tbph]
\centerline{\includegraphics[width=0.75\columnwidth]{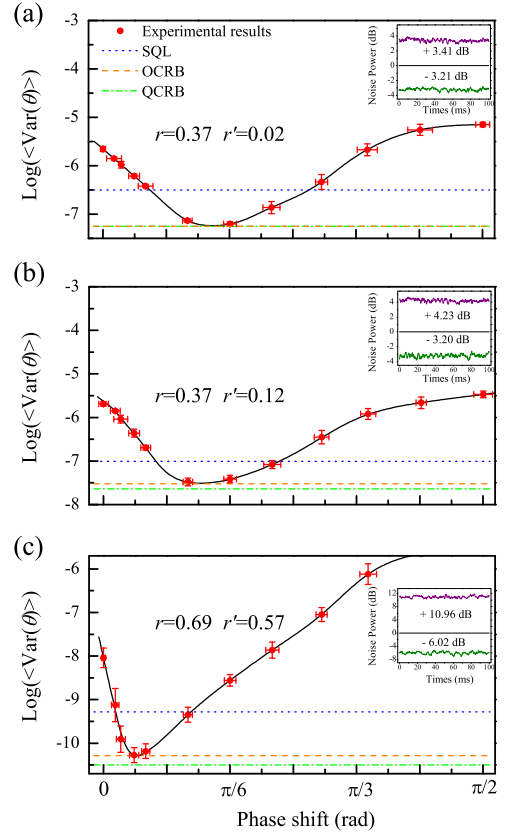}}
\caption{Estimation variance versus phase shift for three different purity
squeezed states as the probe beam. (a).The noise suppression of one
quadrature is measured as - 3.21 dB relative to the SQL while the noise of
the orthogonal quadrature is amplified by + 3.41 dB and the purity of the
probe beam is 0.977. (b). The purity is 0.891 and the noise suppression of
one quadrature is the same as (a) while the noise of the orthogonal
quadrature is increased to + 4.23 dB. (c) The purity is 0.566 and the
measured noise levels are - 6.02 dB and + 10.96 dB. The estimation variances
Var$\left[ \protect\theta \right] $ for twelve phase shifts in the $[0,$ $%
\protect\pi /2]$ range are marked as the red circles with the standard
deviations over 20 repetitions. The dot blue, dash origin and dash dot green
curves correspond to the SQL, OCRB and QCRB, respectively.}
\end{figure}

Then we analyze the effect of the purity of probe beam on the accuracy of
phase estimation. The estimation variances Var$\left[ \theta \right] $ for
squeezed states with different purity are shown in Fig. 4. The estimation
variances measured at twelve different phase shifts in $[0,$ $\pi /2]$ range
are marked as circles in the figure. The dot blue, dash origin and dash dot
green curves correspond to the SQL, the OCRB and the QCRB which is
calculated with the corresponding equations with same mean photon number,
respectively. The noise power spectra of the probe beam at 3 MHz for
different purity squeezed states measured by spectrum analyzers (SA) are
shown in the insert of Fig. 4. It is obvious that the estimation accuracy
attains OCRB only for one specific phase shift $\theta _{\text{opt}}$ and
estimation accuracy beyond the SQL can be realized in a phase interval $%
\Delta \theta $ near the optimal phase shift $\theta _{\text{opt}}$ \cite%
{oliv}. The larger the squeezing parameter $r$, the higher the estimation
precision. Because the mean photon number of probe beam is the main effect
of the precision of quantum phase estimation,the estimation precision can
also be enhanced with the increasing of factor of extra antisqueezing
parameter $r^{\prime }$ at the same $r$. For a squeezed state of same
squeezing level $r$, the mean photon number of squeezed thermal state is
more than that of pure squeezed state because the existence of extra noise $%
r^{\prime }$ can increase the mean photon number of the squeezed state.
Although the extra antisqueezing parameter has a helpful influence on the
absolute precision of phase estimation, the estimation accuracy is further
away from the QCRB with the increase of $r^{\prime }$ due to the extra loss
and phase fluctuation in antisqueezing quadrature. The accuracy of the
quantum phase estimation can be enhanced with much higher squeezing level at
a given fixed mean photon number and squeezed pure state behaves the optimal
resource to reach QCRB, i. e. Heisenberg limit asymptotically.

\begin{figure}[tbph]
\centerline{\includegraphics[width=0.75\columnwidth]{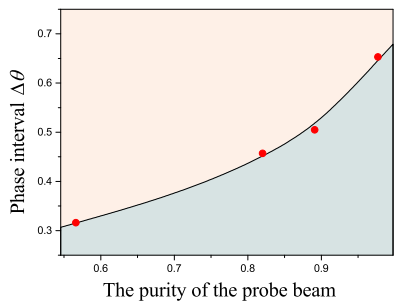}}
\caption{The range of phase interval $\Delta \protect\theta $ for the
estimation variance beyond the SQL versus the purity of the probe beam. The
red circles correspond to the experimental results.}
\end{figure}

Finally, we analyze the influence of purity of probe beam on the phase
interval $\Delta \theta $ of phase estimation beyond the SQL. The range of
quantum phase estimation beyond the SQL for four different purity probe
beams with different squeezing level are shown in Fig. 5. The $\Delta \theta
$ increases with the purity of the probe beam. The range of $\Delta \theta $
is 0.653 at purity of squeezed state with 0.977, which is more than twice of
that at purity of 0.566 without any feedback control.

\section{Conclusion}

Through detailedly analyzing the influence of properties of squeezed thermal
state on the precision of quantum phase estimation, it is found that the
QCRB only can be reached with the help of a squeezed pure state and the
absolute precision of quantum phase estimation can be enhanced with squeezed
state of higher squeezing level. Through controlling the conditions of NOPA,
squeezed states of light with different purity and squeezing level are used
as a probe beam in the experiment of phase estimation. The experimental
results are in good agreement with the theoretical analyses. This provides
us a new direction of simple and convenient phase estimation scheme and it
is also a good reference for the multi-parameter estimation with a
multipartite entanglement.

\section*{acknowledgments}

The authors would like to thank Kaimin Zheng and Shan Ma for helpful
discussions. Our work was supported by the Key Project of the National Key
R\&D program of China (Grant No. 2016YFA0301402), the National Natural
Science Foundation of China (Grants No. 61925503, No. 61775127, No.
11654002, No. 11804246, and No. 11834010), the Program for Sanjin Scholars
of Shanxi Province, and the fund for Shanxi \textquotedblleft 1331
Project\textquotedblright\ Key Subjects Construction.

\end{document}